\documentclass[twocolumn,showpacs,preprintnumbers,superscriptaddress,amsmath,amssymb,prl]{revtex4}       

\usepackage{graphicx}
\usepackage{dcolumn}
\usepackage{bm}
\usepackage{amsmath}
\newcommand{\be}{\begin{equation}}
\newcommand{\ee}{\end{equation}}
\newcommand{\ba}{\begin{eqnarray}}
\newcommand{\ea}{\end{eqnarray}}

\begin{document}

\preprint{APS preprint}

{\bf Comment on ``Analysis of the Spatial Distribution Between
  Successive Earthquakes'' by Davidsen and Paczuski} 

\noindent 

\noindent Davidsen and Paczuski [1] claim to have found evidence
contradicting the 
theory of aftershock zone scaling in favor of scale-free
statistics. We present four elements showing that
Davidsen and Paczuski's results may be insensitive to the existence of
physical length scales associated with aftershock zones or mainshock
rupture lengths, so that their claim is unsubstantiated.

Firstly, the power law exponent $\delta$ of their probability
density distribution (pdf) for distances between pairs of 
successive earthquakes in southern
California is less than $1$. Therefore the
exponent cannot hold for the tail because the pdf cannot be
normalized and hence can at most describe an intermediate asymptotic (if
any). The real tail of the pdf must behave differently, independent of
any finite size scaling.  

Secondly, we performed tests that show the extreme sensitivity of
the suggested power law to the duration of the catalog. We took the
same catalog from the Southern California Earthquake Data Center as in
[1] with the same parameters but contrasted in Figure 1 the highly active
6-month period from June 1, 1992 until December 31, 1992 including
the June 28 M7.3 Landers earthquake (crosses) with the remainder of the
catalog both before and after Landers (circles). Firstly, we see that
removing 6  months of data from a 17 year period causes the power law
to disappear. Secondly, the 
Landers aftershocks show clear signs of scales, such as the bump
marked by an arrow, which may be connected with the
simultaneous aftershocks of the June 28 M$6.4$ Big Bear event and the
July 11 M$5.7$ Mojave earthquake and the rupture length of Landers.

Thirdly, we repeated the same analysis for Japan and northern
California and found no evidence of  robust power laws. For Japan
(Figure 2, circles), we used the JMA catalog 
from January 1984 to 
December 2001 within $(120.0^o E, 150.0^o E)$ by
$(25.0^o N, 45.0^o N)$ 
above magnitude thresholds $m_d=3.0, 3.4 $ and $4.0$ 
resulting in $34163, 13275$ and $6757$ pairs with distances larger
than $2$km, respectively. For northern California (Figure 2, crosses), we
used data in the period from January 1984 until December 2004 (28274
events above magnitude $2.4$). Our results suggest that the power law
found by Davidsen and Pazcuski [1] may not be robust with 
respect to location, catalog and window size. 

Finally, we show in the inset of figure 2 that a model [2] that
explicitely obeys 
aftershock zone scaling can 
reproduce the observed histogram, demonstrating that the statistic may
not be sensitive to the scales. We simulated a seismic
catalog using a 3D version of the ETAS model [2] that explicitely includes
the scale of each mainshock rupture length $l_r(m)=0.02 \times 10^{0.5m}$
in the spatial aftershock decay distribution with distance $d$
according to $P(d) \sim (l_r(m)+d)^{-(1+\mu)}$ and 
initiated the catalog with a Landers-like $M7.3$ mainshock. We used
the parameters ($b=\alpha=1, p=1.1, c=0.0001, k=0.0022, \mu=2$,
background=5.0 per day in a 700km by 700km window). We show 
that the same analysis of data generated with aftershock zone scaling
leads to an apparent power law that shows no sign of the aftershock
zone scale $l_r \sim 90km$.  

\begin{figure}[h!]
\includegraphics[width=7cm]{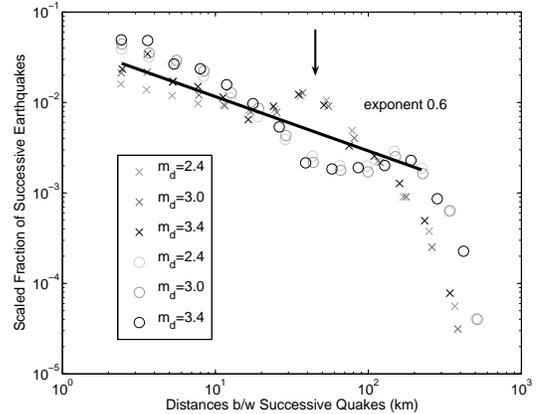}
\caption{\label{Fig1} Distribution of the epicentral distances between
  successive earthquakes in southern California for magnitude
  thresholds $2.4, 3.0, 3.4$ (light to dark markers): 6 month period
  from June through December 1992 (crosses) consisting mainly of
  Landers and its aftershocks contrasted with the remainder of the 17
  year catalog (circles). 
}
\end{figure}

\begin{figure}[h!]
\includegraphics[width=7cm]{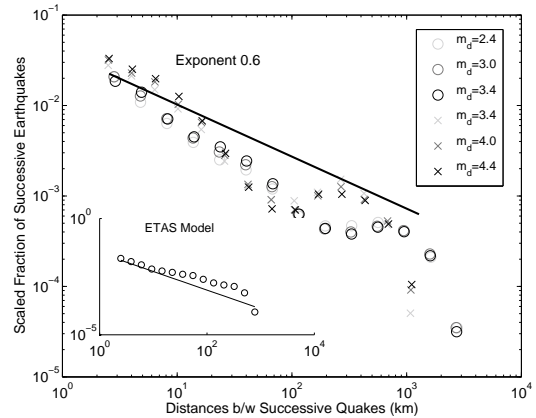}
\caption{\label{Fig2} Distribution of the epicentral distances between
  successive earthquakes in the JMA catalog (circles) from January 1984 until
  December 2000 above magnitude thresholds $3.4$, $4.0$ and $4.4$ and
  in northern California (crosses) from January 1984 until December 2004
  above magnitude thresholds $2.4$, $3.0$ and $3.4$ (data from the
  Northern California Earthquake Data Center). Inset: ETAS model
  simulation.
}
\end{figure}

M.J. Werner $^1$ and D. Sornette$^{2,3}$,\\
$^1$ IGPP, UCLA, and ESS, UCLA, California 90095-1567, USA
$^2$ LPMC, CNRS UMR 6622 and Univ. Nice, 06108 Nice, France
$^3$ MTEC, ETH Zurich, Switzerland

\end{document}